\documentclass[preprintnumbers, nofootinbib, aps, prd, superscriptaddress, 10pt, showkeys, notitlepage]{revtex4-1}

\usepackage{bm}
\usepackage{amsmath}
\usepackage{amsfonts}
\usepackage{graphicx}
\usepackage{amssymb}

\usepackage{epstopdf}
\usepackage{epsfig}
\usepackage{hyperref}

\usepackage{color}

\date{\today}

\def\jnl@style{\it}
\def\aaref@jnl#1{{\jnl@style#1}}

\def\aaref@jnl#1{{\jnl@style#1}}

\def\aj{\aaref@jnl{AJ}}                   
\def\apj{\aaref@jnl{ApJ}}                 
\def\apjl{\aaref@jnl{ApJ}}                
\def\apjs{\aaref@jnl{ApJS}}               
\def\apss{\aaref@jnl{Ap\&SS}}             
\def\aap{\aaref@jnl{A\&A}}                
\def\aapr{\aaref@jnl{A\&A~Rev.}}          
\def\aaps{\aaref@jnl{A\&AS}}              
\def\mnras{\aaref@jnl{Mon.~Not.~Roy.~Astron.~Soc.}}             
\def\prd{\aaref@jnl{Phys.~Rev.~D}}        
\def\prc{\aaref@jnl{Phys.~Rev.~C}}  
\def\prl{\aaref@jnl{Phys.~Rev.~Lett.}}    
\def\qjras{\aaref@jnl{QJRAS}}             
\def\skytel{\aaref@jnl{S\&T}}             
\def\ssr{\aaref@jnl{Space~Sci.~Rev.}}     
\def\zap{\aaref@jnl{ZAp}}                 
\def\nat{\aaref@jnl{Nature}}              
\def\aplett{\aaref@jnl{Astrophys.~Lett.}} 
\def\apspr{\aaref@jnl{Astrophys.~Space~Phys.~Res.}} 
\def\physrep{\aaref@jnl{Phys.~Rep.}}      
\def\physscr{\aaref@jnl{Phys.~Scr}}       
\def\commat{\aaref@jnl{Comm.~Math.~Phys.}}              
\def\science{\aaref@jnl{Science}}               
\def\cqg{\aaref@jnl{Classical Quant.~Grav.}}            
\def\jpcs{\aaref@jnl{JPCS}}                                     
\def\ijmpd{\aaref@jnl{Int.~J.~Mod.~Phys.~D}}                    
\def\grg{\aaref@jnl{Gen.~Relat.~Gravit.}}               
\def\rpp{\aaref@jnl{Rep.~Prog.~Phys.}}          
\def\npa{\aaref@jnl{Nucl.~Phys.~A}}        
\def\lrr{\aaref@jnl{Living Rev.~Rel.}}                   
\def\jcap{\aaref@jnl{J.~Cosmology Astropart.~Phys.}}    
\def\rmp{\aaref@jnl{Rev.~Mod.~Phys.}}   

\begin{document}

\title{Axial Quasi-Normal Modes of Scalarized Neutron Stars with Massive
	Self-Interacting Scalar Field}

\begin{abstract}
	
	We study the axial quasi-normal modes of neutron stars in scalar-tensor theories with massive scalar field including a self-interacting term in the potential. Various realistic equations of state including nuclear, hyperonic and hybrid matter are employed. Although the effect of spontaneous scalarization of neutron stars can be very large, binary pulsar observations and gravitational wave detections significantly constrain the massless
	scalar-tensor theories. If we consider a properly chosen nonzero mass for the scalar field, though, the scalar-tensor parameters cannot be restricted by the
	observations, resulting in large deviation from pure general relativity. With this motivation in mind, we extend the universal relations for axial quasi-normal modes known in general relativity to neutron stars in massive scalar-tensor theories with self-interaction using a wide range of realistic EOS. We confirm the universality of the scaled frequency and damping time in terms of the compactness and scaled moment of inertia for neutron stars with and without massive scalarization.
\end{abstract}

\author{Zahra Altaha Motahar}\email[]{zahra.motahar@uni-oldenburg.de}
\affiliation{Institut f\"ur  Physik, Universit\"at Oldenburg  Postfach 2503,
	D-26111 Oldenburg, Germany}

\author{Jose Luis Bl\'azquez-Salcedo}\email[]{jose.blazquez.salcedo@uni-oldenburg.de}
\affiliation{Institut f\"ur  Physik, Universit\"at Oldenburg  Postfach 2503,
	D-26111 Oldenburg, Germany}

\author{Daniela D. Doneva}\email[]{daniela.doneva@uni-tuebingen.de}
\affiliation{Theoretical Astrophysics, Eberhard Karls University of T\"ubingen, T\"ubingen 72076, Germany}
\affiliation{INRNE - Bulgarian Academy of Sciences, 1784  Sofia, Bulgaria}

\author{Jutta Kunz}\email[]{jutta.kunz@uni-oldenburg.de}
\affiliation{Institut f\"ur  Physik, Universit\"at Oldenburg  Postfach 2503,
	D-26111 Oldenburg, Germany}

\author{Stoytcho S. Yazadjiev}\email[]{yazad@phys.uni-sofia.bg}
\affiliation{Theoretical Astrophysics, Eberhard Karls University of T\"ubingen, T\"ubingen 72076, Germany}
\affiliation{Department of Theoretical Physics, Faculty of Physics, Sofia University, Sofia 1164, Bulgaria}
\affiliation{Institute of Mathematics and Informatics, Bulgarian Academy of Sciences, Acad. G. Bonchev Street 8, Sofia 1113, Bulgaria}

\vspace{0.5truecm}

\vspace{0.5truecm}

\vspace{0.5truecm}
\date{
\today}

\maketitle

\section{Introduction}
Neutron stars represent valuable astrophysical laboratories to investigate various aspects of gravity.
Moreover, neutron stars consist of extremely dense matter at supranuclear densities that is poorly understood due to the fact that it cannot be produced in ground based laboratories. All this makes the astrophysical observations of neutron stars extremely important. 
In this context, the discovery of gravitational waves in particular by the merger of binary neutron stars GW170817 \cite{Abbott2017}, that was also observed as a $\gamma$-ray burst (GRB 170817A), has led to a new era in physics and astronomy, where multi-messenger observations including and triggered by gravitational waves promise to elucidate some of the most intriguing and at the same time cataclysmic events in nature \cite{Abbott2017a}. 

Gravitational wave observations are essential in testing the strong field regime of general relativity (GR). For example, the joint observation of GW170817 in gravitational waves and in the electromagnetic spectrum has already ruled out a large number of alternative theories of gravity \cite{Baker:2017hug,Ezquiaga:2017ekz,Sakstein:2017xjx}. The studies of neutron star mergers in alternative theories of gravity, especially during the inspiral phase, are supposed to give us also invaluable information when confronted against the future gravitational wave observations \cite{Barausse2013,Shibata2014,Taniguchi2015,Sennett2016,Ponce2015,Sennett2017,Sagunski2018,Berti:2018cxi,Berti:2018vdi}. On another front, the detailed study of neutron star quasi-normal modes (QNM), that can be excited for example after binary mergers and core-collapse, in modified gravity would also allow us to set constraints on the deviations from pure GR. The theoretical advance in this field is a timely effort due to the expected large number of gravitational wave detections in the future made by the next operation of LIGO/VIRGO and the interferometers that are either already planned or under construction, such as KAGRA, LIGO in India, and the Einstein Telescope.

One of the alternative theories of gravity that survived after the joint gravitational wave and electromagnetic observation of GW170817, are the scalar-tensor theories (STT). These theories are well posed and pass through all the current observations for a proper choice of the free parameters.  Particularly interesting subclasses of STT are the ones that are perturbatively equivalent to pure GR in the weak field regime and thus fulfill by definition all the weak field observations, while non-perturbative effects can be observed for strong fields such as the spontaneous scalarization of neutron stars \cite{Damour1993,Harada1998,Salgado1998}. Scalarized slowly rotating neutron stars were considered in \cite{Damour1996,Sotani2012,Pani2014a,Motahar:2017blm} while the rapidly rotating case was examined in \cite{Doneva2016}. The recent observations of massive neutron stars in binary systems \cite{Demorest10,Freire2012}, though, have severely limited the range of allowed values of the parameters, leaving a possibility for larger deviations from pure GR only in the rapidly rotating case. All of these studied were done in the case of a massless scalar field. If we consider a nonzero mass for the scalar field, the dipole gravitational radiation will be suppressed at length scales of the  order of the scalar field Compton wavelength. This will reconcile the theory with the binary pulsar observations for a wide range of parameters resulting in the possibility for very large deviations from pure GR. Neutron stars in such theories were first constructed in \cite{Popchev2015,Ramazanoglu2016,Yazadjiev2016} including the cases of both static and rapidly rotating models. Furthermore, it was shown that adding a self-interacting term in the potential can lead to interesting effects and reconcile the theory with the observations for an even larger set of parameters \cite{Staykov2018}.

It is natural to attempt to further study the astrophysical implications of scalarized neutron stars putting emphasis on the connection with the future gravitational wave observations. As we commented above, one of the possible channels of emitting gravitational radiation from such objects are the QNMs of neutron stars. One of the important challenges in this direction is to find a way to connect the observed gravitational wave frequencies and damping times to the properties of the emitting neutron star in order to obtain information for the high density nuclear matter EOS, the underlying gravitational theory and the possible deviations from pure GR.  This is a difficult task on its own due to the large uncertainty in the EOS and the fact that often there is a degeneracy between effects coming from modifications of Einstein's theory of gravity and the EOS uncertainty. An elegant way out of this problem is to construct equation of state independent relations \cite{Yagi:2016bkt} between the QNM frequencies and damping times on the one hand and the neutron star properties (and the free parameters of the alternative theories of gravity) on the other \cite{Andersson96,Andersson98a,Tsui2005}. Such relations were constructed in several alternative theories of gravity, both for the polar and the axial QNMs \cite{Sotani04,Sotani2005,Staykov2015,Silva2014,Yazadjiev:2017vpg,Blazquez-Salcedo:2015ets,Blazquez-Salcedo2018a,Blazquez-Salcedo2018,AltahaMotahar2018}, and can be potentially used to constrain the strong field regime of gravity.

In this work we extend our previous studies of QNMs in STT to the case of massive STT with self-interaction  employing various realistic EOS, including pure nuclear matter, nuclear matter with hyperons, hybrid nuclear and quark matter, and hadron with quark matter. More precisely, we focus on the axial quasinormal modes of static and spherically symmetric massive scalarized neutron stars and compare the results with those of massless STT and GR.  In addition, we extend the universal relations for axial QNMs known in general relativity to such scalarized neutron stars.

In section \ref{sec:Basic Eqs} we present the mathematical and physical framework of our neutron star model including the basic equations for axial QNMs of static neutron stars in massive STT. A short description of the employed realistic EOSs and the numerical method in our model are also included in section \ref{sec:Basic Eqs}. In section \ref{sec:Results}, our results for the massive scalarized neutron star models, as well as the GR solutions are given. 
We illustrate the universality of some matter-independent relations between scaled parameters of the axial QNMs in this section.
Finally, we summarize our study and present some possible outlook in section \ref{sec:Conclusions}.

\section{The model}\label{sec:Basic Eqs}

\subsection{Massive scalar-tensor theory with self-interaction}\label{sec:STT}
The general form of the  Einstein frame action of STT is given by

\begin{equation}
S= \frac{1}{16\pi G_*}\int d^4x \sqrt{-g} \left[{\cal R} -
2g^{\mu\nu}\partial_{\mu}\varphi \partial_{\nu}\varphi- V(\varphi)\right]+ S_{m}[\Psi_{m}; \mathrm{A}^{2}(\varphi)g_{\mu\nu}] ,
\label{action_Einstein}
 \end{equation}
where $G_*$ is the bare gravitational constant, ${\cal R}$ is the Ricci scalar with respect to the Einstein frame metric ${g_{\mu\nu}}$, and $\varphi$ is the scalar field with potential $V(\varphi)$. The second term denoted by $S_m$ is the action of the additional matter fields $\Psi_{m}$. Since we are working in the Einstein frame, a direct coupling between the matter and the scalar field appears through the function $A(\varphi)$. One should point out that such a coupling is present only in the Einstein frame while in the physical Jordan frame the scalar field influences the matter only through the spacetime metric in order to not violate the weak equivalence principle. The function $A(\varphi)$ also controls the transformation between the Einstein and the Jordan frame, i.e. ${\tilde g}_{\mu\nu} = {A}^{2}(\varphi)g_{\mu\nu}$ where ${\tilde g}_{\mu\nu}$ is the Jordan frame metric.
In the present paper we will work with a non-negative scalar field potential with self-interaction. The natural and the simplest choice is the following $Z_2$ symmetric potential
\begin{equation}
V(\varphi)=2m^2_{\varphi}\varphi^2 + \lambda \varphi^4,
\end{equation}
where $m_{\varphi}$ is the scalar-field mass  and $\lambda\ge 0$ is a positive parameter with dimension of $length^{-2}$ \cite{Popchev2015,Ramazanoglu2016,Yazadjiev2016,Doneva2016}.

The Einstein frame field equations derived by variation of the action (\ref{action_Einstein}) with respect to the metric and the scalar field, have the following form
\begin{eqnarray} 
{\cal R}_{\mu\nu} - \frac{1}{2}g_{\mu\nu}{\cal R} &=&
  2\partial_{\mu}\varphi \partial_{\nu}\varphi   -
g_{\mu\nu}g^{\alpha\beta}\partial_{\alpha}\varphi
\partial_{\beta}\varphi
+ 8\pi T_{\mu\nu} , \label{eq:FieldEq1} \\ \notag \\
 \nabla^{\mu}\nabla_{\mu}\varphi &=& - 4\pi k(\varphi)T, \label{eq:FieldEq2}
\end{eqnarray}
where ${\cal R}_{\mu\nu}$ is the Ricci tensor,  $T_{\mu\nu}$ is the stress-energy tensor with a trace $T$, and $k(\varphi)= \frac{d\ln({A}(\varphi))} {d\varphi}$  is the logarithmic derivative of the function $A(\varphi)$. 

In the present paper we will consider the following standard form of the coupling function
\begin{equation}
{A}(\varphi)=e^{\frac{1}{2}\beta\varphi^2} \ , \ \ \ k(\varphi)=\beta \varphi
\label{A} 
\end{equation}
that leads to a STT that is indistinguishable from GR in the weak field regime while non-perturbative effects such as spontaneous scalarization exist for strong fields.

We model the neutron stars as a self-gravitating perfect fluid. Thus the Einstein frame stress energy momentum tensor ${T}_{\mu\nu}$ is given by 
\begin{equation}
 T_{\mu\nu}= (\varepsilon + p)u_{\mu} u_{\nu} + p g_{\mu\nu} , 
\end{equation}
where $\varepsilon$, $p$ and $u$ represent the energy density,
the pressure and the four-velocity of the fluid, respectively. 

Since we will be working with a static spherically symmetric neutron star model, the Einstein frame metric takes the following general form
\begin{equation}
ds^{2}=-e^{2\nu(r)}dt^{2}+e^{2\psi(r)}d r^{2} + r^{2} (d \theta^{2} 
+\sin^{2}\theta d \phi^{2}),
\label{eq:metric}
\end{equation}
where $\nu$ and $\psi$ are the metric functions depending only on the radial coordinate $r$. The dimensionally reduced field equations for these metric functions and the scalar field that follow from \ref{eq:FieldEq1} and \ref{eq:FieldEq2} can be found in \cite{Yazadjiev2016}.
 
All of the above equations were in the Einstein frame that is chosen for convenience. Of course the final quantities that are presented in the results section are transformed to the physical Jordan frame. The transformations between the two frames is commented in detail in \cite{Yazadjiev2016,Doneva2016}.  

We will use $c=G_*=1$ units for simplicity.

\subsection{QNMs: general formalism and axial perturbations}

In order to study axial perturbations, we follow the same procedure as in  \cite{Blazquez-Salcedo2018,AltahaMotahar2018}. Since we are considering axial QNMs of neutron stars, the perturbations of the scalar field, the pressure and the energy density vanish since they transform as scalars under reflections of the angular coordinates. The axial perturbation of the metric takes the following form:
\begin{eqnarray}
H^{axial}_{\mu\nu} = \left(
\begin{array}{cccc}
0 & 0 &  h_{0}(t,r) S^{lm}_{\theta}(\theta,\phi) & h_{0}(t,r)S^{lm}_{\phi}(\theta,\phi) \\
0 & 0 &  h_{1}(t,r)S^{lm}_{\theta}(\theta,\phi)& h_{1}(t,r)S^{lm}_{\phi}(\theta,\phi) \\
h_{0}(t,r)S^{lm}_{\theta}(\theta,\phi) & h_1(t,r)S^{lm}_{\theta}(\theta,\phi) & 0 & 0 \\
h_{0}(t,r)S^{lm}_{\phi}(\theta,\phi) & h_1(t,r)S^{lm}_{\phi}(\theta,\phi)  & 0 & 0  \\
\end{array}
\right),
\end{eqnarray}  
where $(S^{lm}_{\theta}(\theta,\phi),S^{lm}_{\phi}(\theta,\phi))=(-\partial_{\phi}Y_{lm}(\theta,\phi)/\sin\theta,\,\sin\theta \partial_{\theta}Y_{lm}(\theta,\phi))$  and the spherical harmonics are denoted by $Y_{lm}(\theta,\phi)$. It is possible to define a single master equation governing the axial peturbations, by introducing the variable $X=\frac{h_1 e^{\psi-\nu }}{r}$. Thus one can obtain
\begin{equation}
\frac{\partial^2 X}{\partial t^2} 
-e^{\nu-\psi} \frac{\partial}{\partial r}\left[e^{\nu-\psi}\dfrac{\partial X}{\partial r}\right]
+ e^{2\nu}\left[ \dfrac{l(l+1)}{r^{2}}-\dfrac{3}{r^{2}}(1-e^{-2\psi})
+ 4\pi {\cal A}^{4}(\varphi)(\tilde{\varepsilon}-\tilde{p})+\frac{1}{2} V(\varphi)\right] X =0
\end{equation}
where $\tilde{\varepsilon}$ and $\tilde{p}$ are the Jordan frame energy density and pressure connected to the Einstein frame ones as $\varepsilon = {\cal A}^{4}(\varphi)\tilde{\varepsilon}$ and $p = {\cal A}^{4}(\varphi)\tilde{p}$. The above equation is the generalized Regge-Wheeler equation for massive STT neutron stars. The pure general relativistic version of the  Regge-Wheeler equation  is obtained in the limit of zero scalar field $\varphi=0$, while the equation for axial perturbations of static STT black holes is recovered if we consider zero matter content $\tilde{p}=\tilde{\varepsilon}=0$.

We can cast the above time-dependent equation in a Schr\"odinger type equation by assuming a  time dependence of the form $X(r,t)=X(r)e^{-i\omega t}$:  
\begin{equation}
 -e^{\nu-\psi} \frac{\partial}{\partial r}\left[e^{\nu-\psi}\dfrac{\partial X}{\partial r}\right]
+\left(\omega^{2}+ e^{2\nu}\left[ \dfrac{l(l+1)}{r^{2}}-\dfrac{3}{r^{2}}(1-e^{-2\psi})
+ 4\pi {\cal A}^{4}(\tilde{\varepsilon}-\tilde{p})+\frac{1}{2} V(\varphi)\right]\right) X =0.
\label{eq:d2X}
\end{equation}
Here the QNM frequency $\omega$ is a complex variable, where the real part $\omega_R$ controls the oscillations while the imaginary part $\omega_I$ is connected to the damping time of the modes. 

The boundary conditions one has to impose in order to calculate QNMs of neutron stars are the following. The perturbations should be regular at the center of the star which is satisfied when  
\begin{equation}
X (r\rightarrow 0) \sim r^{l+1}.
\end{equation}
At infinity the general solution for the perturbation $X(r)$ is an admixture of an ingoing and outgoing wave:
\begin{eqnarray}
X \sim A_{in} e^{i\omega (t+R)} + A_{out} e^{i\omega (t-R)},
\end{eqnarray}
where $R$ is the tortoise coordinate defined via $dR=e^{\psi-\nu} dr$. In order to obtain the QNM frequencies the ingoing wave contribution should be zero leaving $X(r)$ in the form of a purely outgoing wave. 

\subsection{Numerical method and equation of state}

In order to obtain the background configurations, we integrate the ordinary differential equations for static and spherically symmetric neutron stars in a compactified coordinate $x = r/(r+r_s)$, where $r_s$ is the coordinate radius of the star with $p(r=r_s)=0$. In this way, we impose the regularity and asymptotic flatness conditions at $x=0$ and $x=1$, respectively. The surface of the neutron star is located at $x=1/2$ with $p(x=1/2)=0$. The ODEs are integrated with the package COLSYS \cite{Ascher:1979iha}.

In order to obtain the quasinormal modes of these configurations, we follow the method previously described in \cite{BlazquezSalcedo:2012pd,Blazquez-Salcedo:2013jka,Blazquez-Salcedo:2015ets,Blazquez-Salcedo2018,AltahaMotahar2018}. In a summary, instead of solving the full perturbation equation (\ref{eq:d2X}), we integrate the related Riccati type equation for the phase function $g = \frac{1}{X}\frac{dX}{dr}$. Two solutions are then generated. One of them extends from the center of the star, where regularity of the perturbation is imposed, up to a some point outside of the star, $x_i>1/2$. The second solution is obtained outside of the star and using exterior complex scaling, an outgoing wave behaviour is imposed at infinity. The QNMs are obtained when both solutions are continuous at $x=x_i$, and the value of $x_i\sim 0.6-0.7$ can be adjusted to improve the precision. This method has been successfully used in other settings \cite{PhysRevA.44.3060,BlazquezSalcedo:2012pd,Blazquez-Salcedo:2013jka,Blazquez-Salcedo:2015ets,Blazquez-Salcedo2018,AltahaMotahar2018} (for a review see \cite{Blazquez-Salcedo:2018pxo}), and allows to obtain the QNMs with a precision better than $10^{-3}$.

Concerning the equation of state, in this paper we investigate a total of 13 realistic models plus a simple relativistic polytrope for comparison.
All the EOSs considered are consistent with the current $2 M_{\odot}$ bound for the maximum mass observed in neutron star candidate pulsars (PSR J1614-2230 \cite{Demorest:2010bx} and PSR J0348+0432 \cite{Antoniadis:2013pzd}) (also see \cite{Lattimer:2013hma,Ozel:2016oaf,Most:2018hfd} for a discussion of other constraints on the mass and radius).

Following the nomenclature used in previous works \cite{Blazquez-Salcedo:2015ets,Blazquez-Salcedo2018,AltahaMotahar2018}, we study: 

\begin{itemize}
	\item 2 standard nuclear matter EOS:  
	SLy \cite{Douchin2001} and APR4 \cite{AkmalPR};
	\item 5 EOS including hyperon matter:  
	BHZBM \cite{Bednarek:2011gd}, GNH3 \cite{Glendenning:1984jr}, 
	H4 \cite{Lackey:2005tk} and WCS1-2 \cite{Weissenborn:2011ut}.
	\item 3 EOS with hybrid quark+nuclear matter:
	ALF2-4 \cite{Alford:2004pf}, 
	and WSPHS3 \cite{Weissenborn:2011qu}.
\end{itemize}

In addition, we study a new family of EOS recently considered in \cite{Paschalidis:2017qmb}.
These new equations include hybrid hadron+quark matter at the core of the stars, and allow for the existence of additional branches of stable neutron stars. In this paper we analyze 3 different cases as examples: ACS-I \cite{PhysRevLett.119.161104} with $j=0.43$, fixing $\tilde p_{tr}=1.7 \times 10^{35} $ $ dyn $ $ cm^{-2} $, $\tilde \varepsilon_{tr}=8.34 \times 10^{14} $ $ g $ $ cm^{-3} $ and $ C_{s}^{2} = 0.8$; ACS-II \cite{PhysRevC.96.045809} with $j=0.8$ and $j=1.0$, fixing $\tilde p_{tr}=8.34 \times 10^{34} $ $ dyn $ $ cm^{-2} $, $\tilde \varepsilon_{tr}=6.58 \times 10^{14}$ $ g $ $ cm^{-3} $, 
and $ C_{s}^{2} = 1$.
Finally, the polytropic EOS is described by the equation $\tilde \varepsilon = \frac{\tilde p}{\Gamma - 1}+(\frac{\tilde p}{K})^{\Gamma}$, with $K =1186$ and $\Gamma=2.34$.

In order to implement the EOSs in our numerical code, we make use of the piecewise polytropic interpolation 
presented in \cite{Read:2008iy} for ALF2-4, APR4, GNH3, H4 and SLy. For the remaining EOSs, which are available in tabulated format, we use a piecewise monotonic cubic Hermite interpolation.

\section{Results} \label{sec:Results}

\subsection{QNM frequencies and damping times}  \label{sec:Models}

\begin{figure}[t!]
	\includegraphics[width=.67\textwidth, angle =-90]{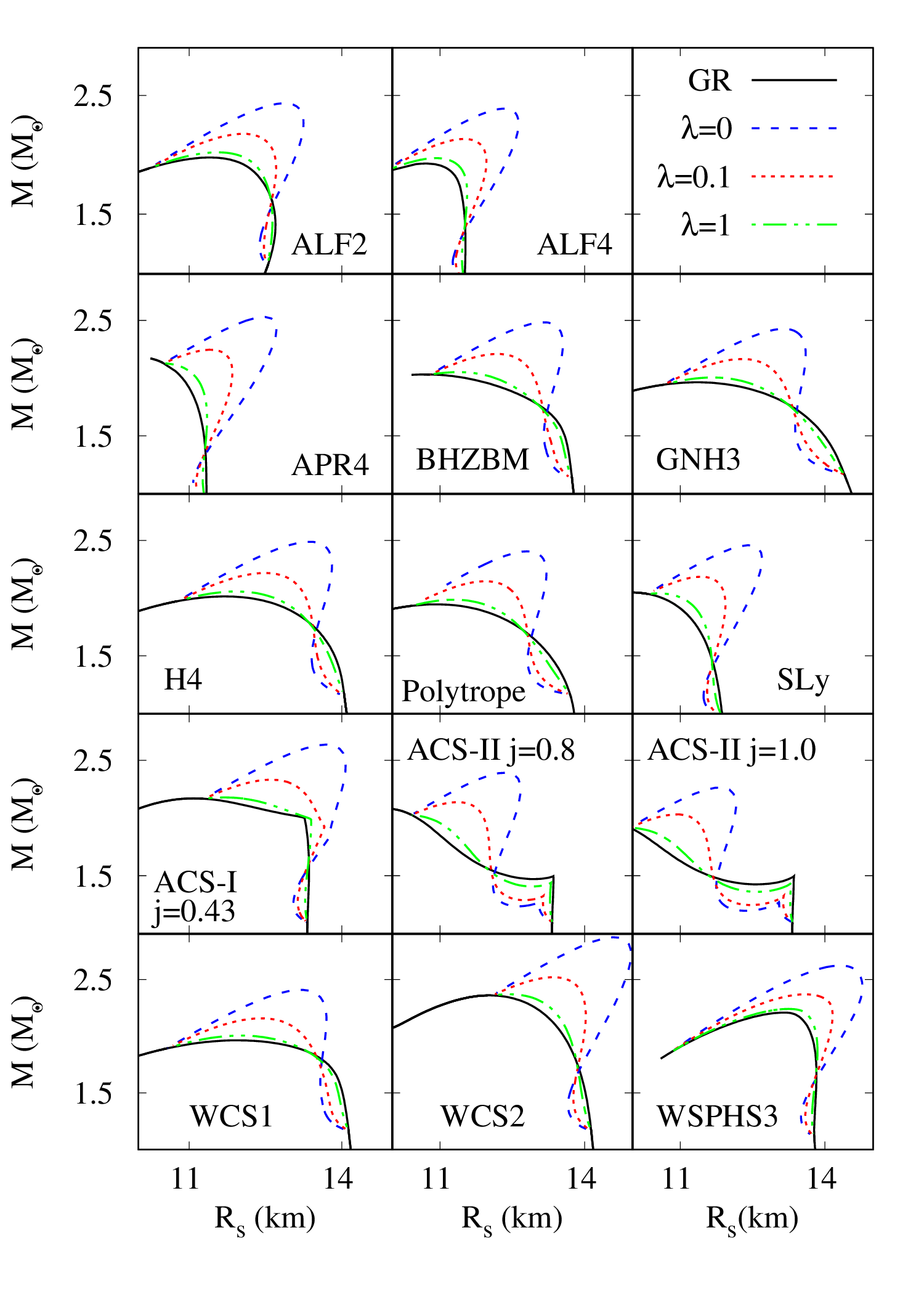}
	\caption{
		Total mass $M$ (in solar masses $M_\odot$) 
		versus the physical radius $R_s$ (in km) of the neutron star models considered. Each panel corresponds to a different EOS, as indicated by the label. The black solid curve corresponds the GR (non-scalarized) configurations. The color curves correspond to scalarized stars with $A=e^{\frac{1}{2}\beta\varphi^2}$, $\beta=-6$, mass of scalar field $m_{\varphi}=10^{-3}$, and different values of the self-interaction parameter $\lambda$: dashed blue curve for $\lambda=0$ (no self-interaction), dotted red curve for $\lambda=0.1$, and double-dashed green curve for $\lambda=1$.
	}
	\label{plot_M_R_multi}
\end{figure}

Let us start by presenting the static neutron star models. We show the total mass $M$ (in solar masses $M_\odot$)
versus the physical (Jordan frame) radius $R_s=A(\varphi_s)r_s$ (in km) 
in Fig.~\ref{plot_M_R_multi} for the sequences of background models that will be perturbed later in order to obtain the QNM spectrum. In each panel we present the mass-radius relation for each particular EOS. With solid black curves we present the GR configurations, and with colored dashed curves the scalarized ones. As one can see from this figure, the chosen set of EOSs covers a large range of stiffness. 

Even though a very large range of values is observationally allowed for the parameter $\beta$ when nonzero mass of the scalar field is considered, the scalarized neutron star models here have been computed for a more moderate and conservative value of the coupling constant: $\beta=-6$. The mass of the scalar field is fixed to $m_{\varphi}=10^{-3}$. The motivation for this choice is the following. Such a value of $m_\varphi$ leads to scalarized neutron star models where the scalar field is exponentially suppressed at a small enough distance from the star, so that the binary pulsar observations cannot impose constraints on the parameter $\beta$ (for a detailed discussion of the observational constraints in the case of a massive scalar field we refer the reader to \cite{Ramazanoglu2016,Yazadjiev2016}). Still, for $m_{\varphi}=10^{-3}$ and $\beta=-6$ considerable deviations from pure GR can be observed.

In general the effect of the massive scalar field is to increase the total mass of the configuration. The largest deviation from GR is obtained for the $\lambda=0$ case without self-interaction (dashed blue curves). Increasing $\lambda$ raises the magnitude of the self-interaction, which makes the deviation from GR smaller (see the dashed red curves corresponding to $\lambda=0.1$). For large enough values of $\lambda$, the difference is so small that the deviation can be hardly tested via observations (see the dotted green curves corresponding to $\lambda=1$, very close to the GR curves). Since the spontaneous scalarization is controlled by $\beta$, and in these figures this is fixed to $\beta=-6$, the three curves of scalarized stars branch off from the same GR configurations.
 
We note that for the EOSs ACS-I j=0.43, ACS-II j=0.8, and ACS-II j=1.0, the mass-radius curves develop a cusp around $R_s = 13.2$ km. This cusp marks the appearance of a quark core inside the star, due to a phase transition in the EOS at high enough densities. This cusp influences other properties of the star such as the QNM spectrum.
 
\begin{figure}
	\centering
\includegraphics[width=.67\textwidth, angle =-90]{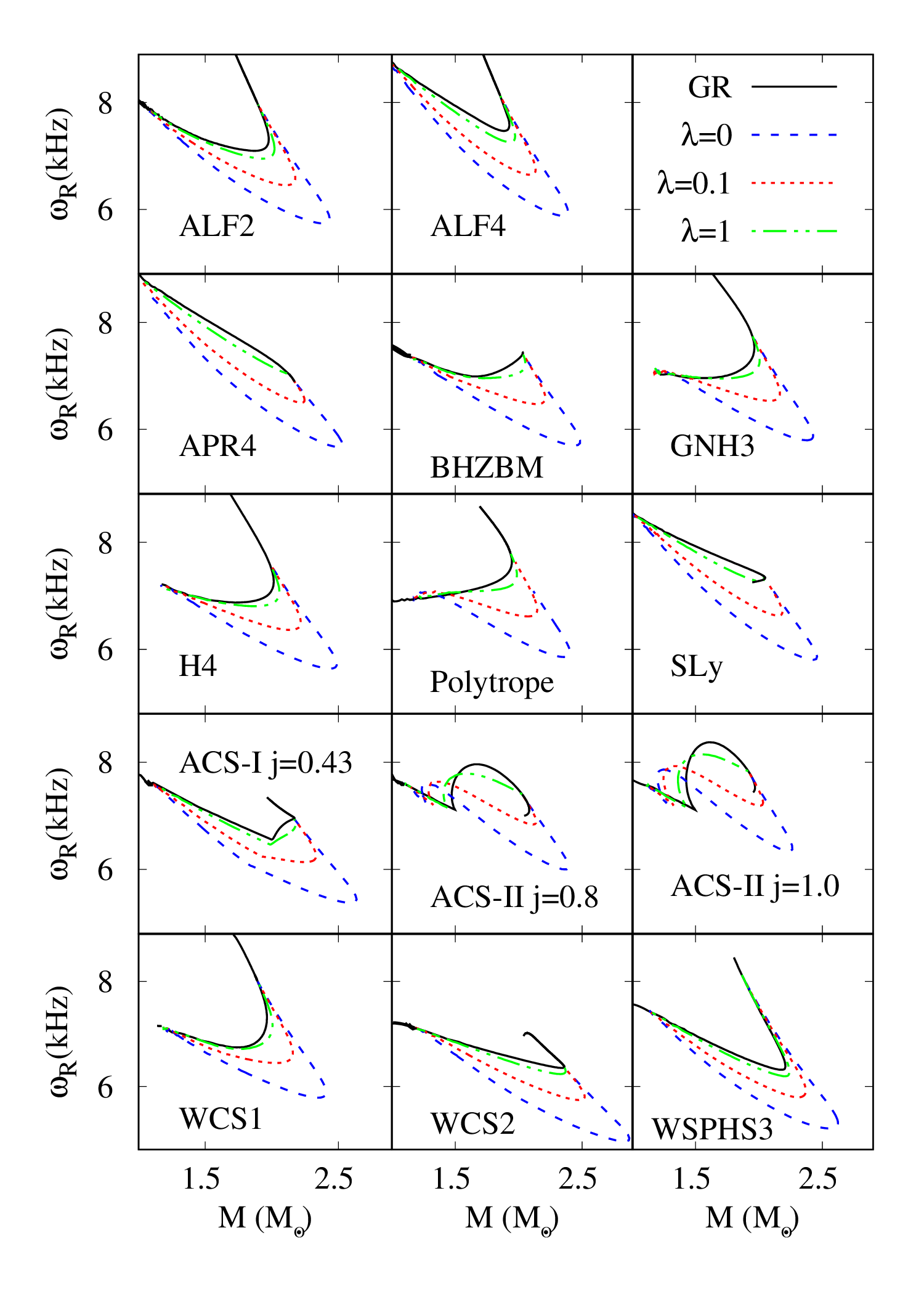}
\includegraphics[width=.67\textwidth, angle =-90]{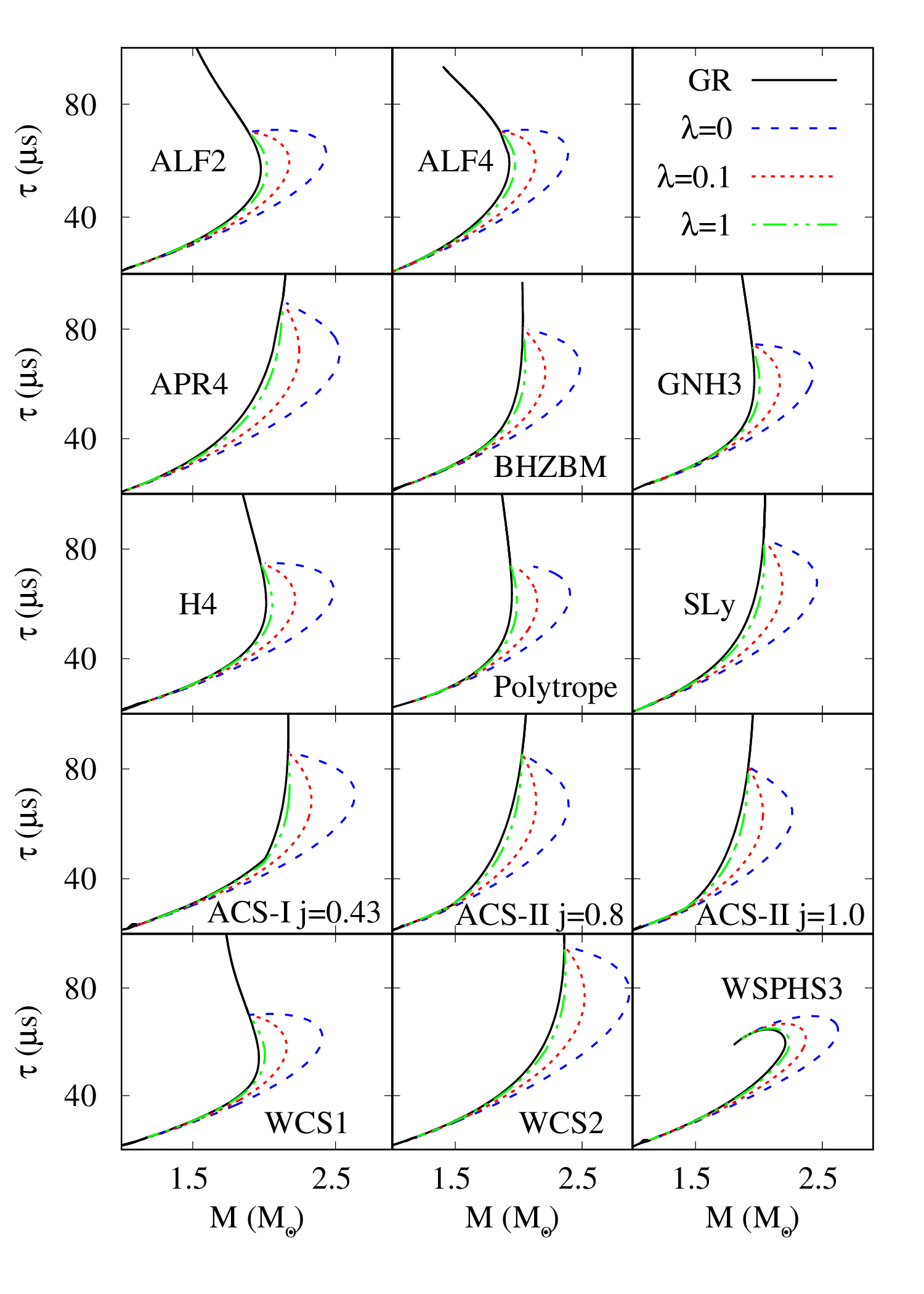}
	\caption{(left) Frequency $\omega_{R}$ (in kHz) of the $l=2$ fundamental wI mode, as a function of the total mass $M$ of the star (in solar masses $M_\odot$). Each panel corresponds to a different EOS, as indicated by the label. The black solid curve corresponds the GR (non-scalarized) configurations. The color curves correspond to scalarized stars with $A=e^{\frac{1}{2}\beta\varphi^2}$, $\beta=-6$, mass of scalar field $m_{\varphi}=10^{-3}$, and different values of the self-interaction parameter $\lambda$: dashed blue curve for $\lambda=0$ (no self-interaction), dotted red curve for $\lambda=0.1$, and double-dashed green curve for $\lambda=1$. (left) Similar figures but for the damping time $\tau$ (in $\mu$s).  
	}
	\label{plot_omegaR_tau_M_multi}
\end{figure} 

For all these neutron star models we study the fundamental $l=2$ w-mode, for which the QNM frequencies $\omega$ are calculated. In Fig.~\ref{plot_omegaR_tau_M_multi}(left) we show 
the oscillation frequency $\omega_R$ (in kHz) versus the total mass $M$ (in solar masses $M_\odot$), following the same notation as in Fig.~\ref{plot_M_R_multi}.

In the figure we can see that the generic effect of the scalarization is to reduce the frequency of the mode. While in GR the stars have a typical frequency around $7-8$ kHz (see the solid black curves in Fig.~\ref{plot_omegaR_tau_M_multi}(left)), the frequencies of the scalarized stars without self-interaction can go below $6$ kHz. Again, increasing the magnitude of the self-interaction reduces the deviation from GR, as can be seen in the $\lambda=0.1$ case (dashed red curves), and specially in the $\lambda=1$ case (dotted green curves). There are some exceptions to this behaviour, in particular, for low values of the total mass of the star, where the frequency of the scalarized stars can slightly increase with respect to the GR case (see for instance the behaviour of the curves for the ACS-II EOS).

We present similar plots in Fig.~\ref{plot_omegaR_tau_M_multi}(right), where we show the damping time $\tau=1/\omega_I$ (in $\mu {\rm s}$) versus the total mass $M$ (in solar masses $M_\odot$). In general, we can see that the more massive scalarized stars do not possess values very different from the unscalarized ones, and the range of values of $\tau$ for the scalarized stars does not depend much on the value of the self-interaction parameter $\lambda$ (typical damping times vary between $30$ and $90$ $\mu$s, as in GR). In other words, in these plots the main deviation of the scalarized curves from the GR curves is due to the increase of the total mass of the star.

We should keep in mind that we have chosen a moderate value of coupling parameter $\beta=-6$. As we commented previously, a scalar field mass $m_{\varphi}=10^{-3}$ allows for much smaller values of $\beta$, for which the deviations from GR can reach much larger values.

\subsection{QNMs universal relation} \label{sec:Universal}

In this section we will study universal relations involving the neutron star QNMs and several global quantities, such as the total mass, compactness and moment of inertia. Universal relations of this type have been studied extensively in the literature (for a recent review see \cite{Doneva:2017jop}). In principle these relations remove to a large extent the dependence of the QNM spectrum on the particular matter composition of the star. Hence, we can use these (almost-)universal relations in order to understand how the effect of changing the theory (in our case, changing the self-interaction parameter $\lambda$) competes with the effect of changing the EOS of the neutron star.

\begin{figure}
	\centering
		\includegraphics[width=.52\textwidth, angle =-90]{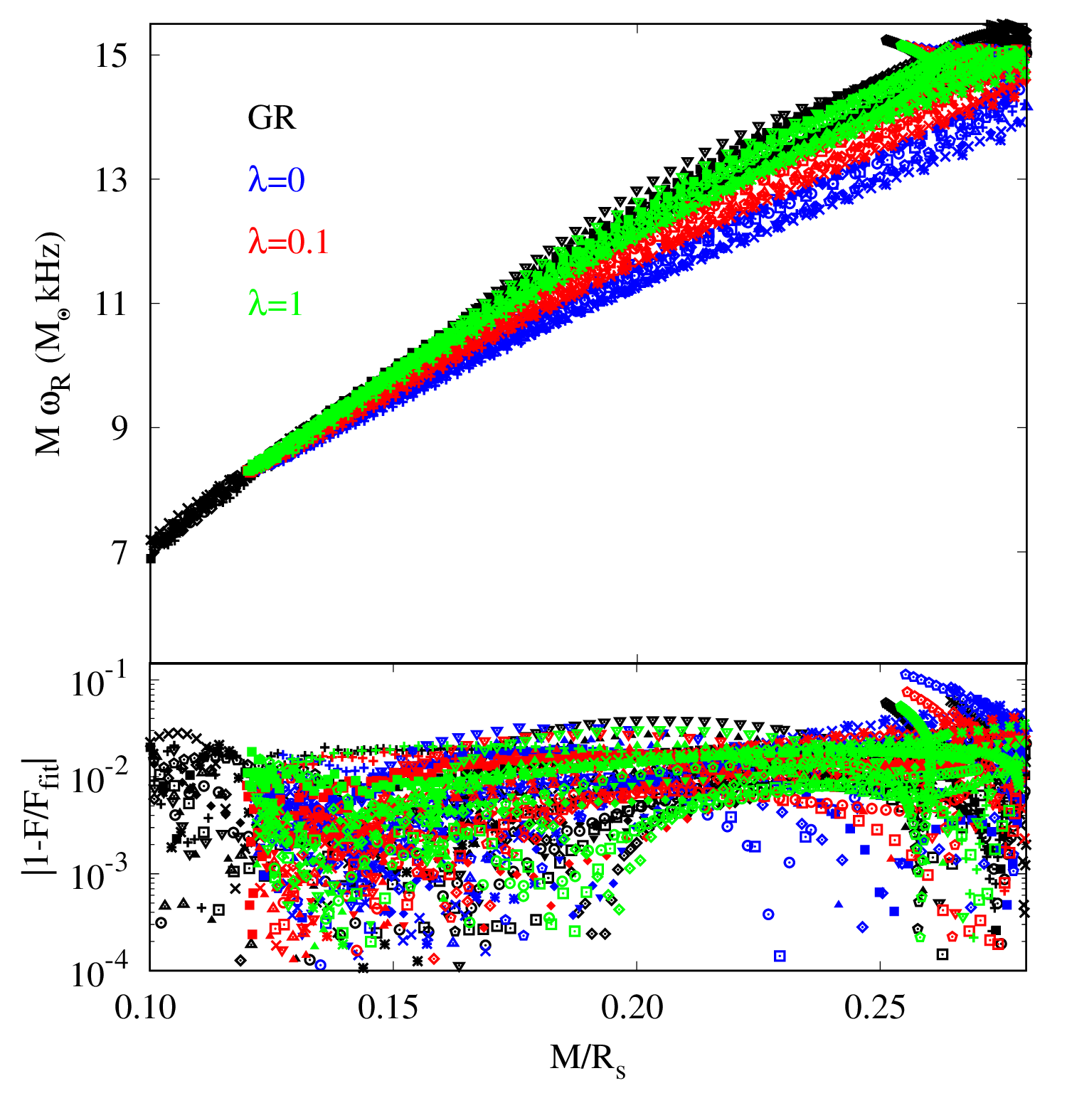}
		\includegraphics[width=.52\textwidth, angle =-90]{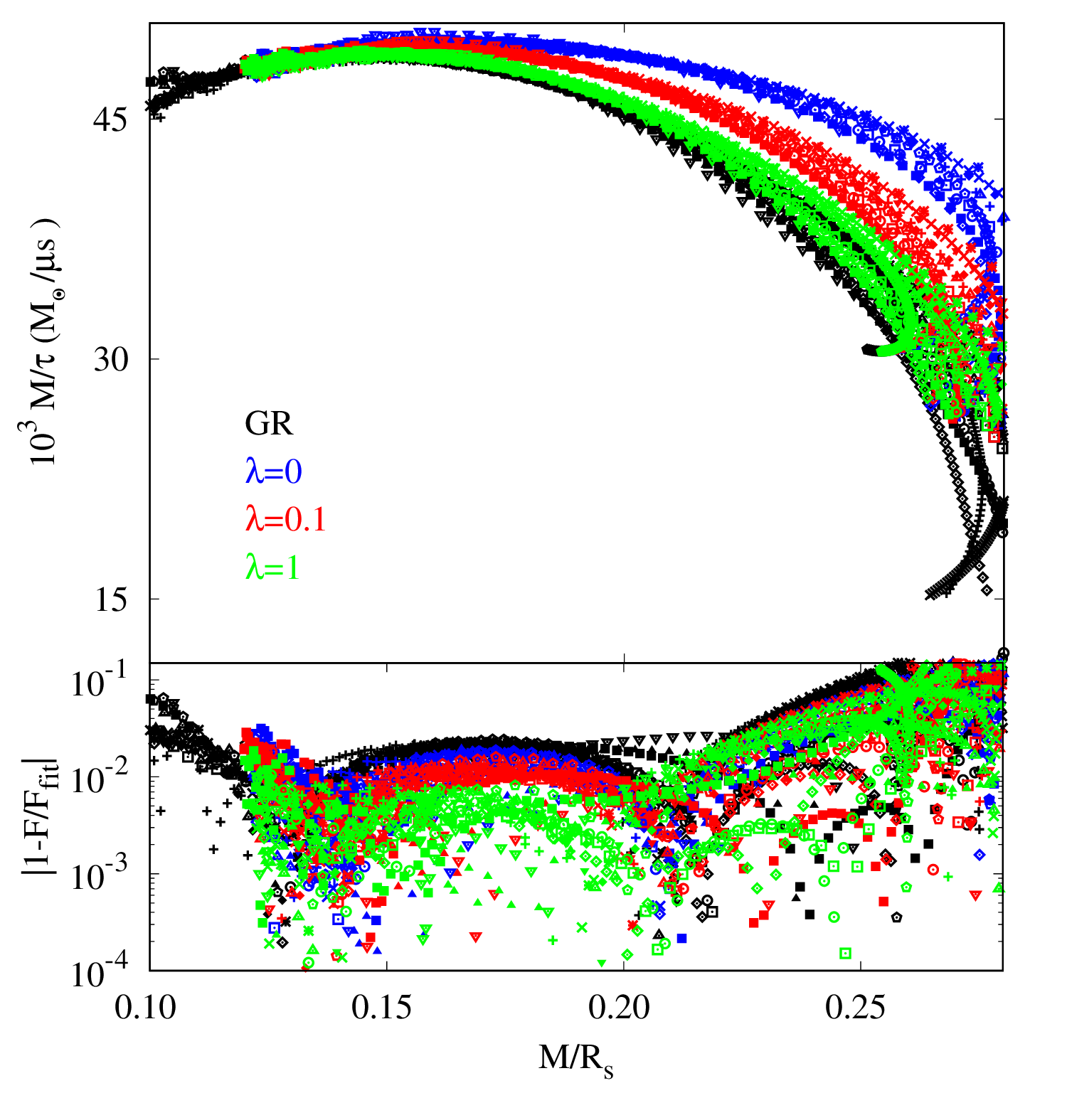}
	\caption{
		(left) Frequency $\omega_{R}$ in kHz scaled to total mass in $M_\odot$ as a function of the compactness $M/R_s$ (fundamental $l=2$ wI mode). In black we show the GR configurations. In colors we show the data for scalarized stars with $A=e^{\frac{1}{2}\beta\varphi^2}$, $\beta=-6$, $m_{\varphi}=10^{-3}$ and different values of $\lambda$: in blue $\lambda=0$, in red $\lambda=0.1$ and in green $\lambda=0.1$. 
		The lower panel shows the relative difference $|1 - F/F_{\rm fit}|$ between the scaled data and the fitting relations presented in eq. (\ref{empiricalfrec}) and Table \ref{tab:universal fits}. 
		(right) A similar figure but showing the inverse of the damping time $\tau$ in $\mu$s scaled to the total mass in $M_\odot$ as a function of the compactness $M/R_s$. 
	}
	\label{plot_MR_omegaR_scaled_v3_w_error1}
\end{figure}

Let us start by some relations involving the total mass of the star and the compactness $\frac{M}{R_s}$. 
In Fig.~\ref{plot_MR_omegaR_scaled_v3_w_error1}(left) we show the frequency of the fundamental $l=2$ wI mode, scaled with the total mass $M \omega_{R}$, as a function of the compactness $M/R_s$. 
With black points we show the data from the GR configurations without scalarization. The remaining colored points show the data for scalarized stars, again with $A=e^{\frac{1}{2}\beta\varphi^2}$, $\beta=-6$, $m_{\varphi}=10^{-3}$ and three different values of $\lambda$: in blue for $\lambda=0$ (no self-interaction), in red for $\lambda=0.1$ and in green for $\lambda=1$. In this figure we have included all the realistic EOSs studied in the previous section (except that we leave out the polytrope, since it behaves slightly differently at low compactness). We can see that when plotted in this way, all the configurations of the same theory tend to follow the same behaviour. The scalarized configurations only deviate slightly from the scalarized GR data (the strongest deviations being found in the $\lambda=0$ scalarized stars). 

We can now consider the data with respect to some simple phenomenological relations. In the present case a quadratic function is enough to describe the relation between the scaled frequency and the compactness,

\begin{eqnarray}
\omega(kHz) = 
\frac{1}{M(M_\odot)}\left[a_1\left(\frac{M}{R_s}\right)^{2} + b_1\frac{M}{R_s} + c_1\right].
\label{empiricalfrec}
\end{eqnarray} 

We fit the data to this formula in the four different cases: GR, $\lambda=0$, $\lambda=0.1$ and  $\lambda=1$. The resulting numerical coefficients are shown in Table \ref{tab:universal fits} of the Appendix. Coming back to Fig.~\ref{plot_MR_omegaR_scaled_v3_w_error1}(left), in the lower panel we show the relative difference $|1 - F/F_{\rm fit}|$ between the scaled data and the resulting fits with equation (\ref{empiricalfrec}). We can see that the typical deviation of the data from the empirical relation is order $10^{-2}$. 

Similarly, in Fig.~\ref{plot_MR_omegaR_scaled_v3_w_error1}(right) we show the inverse of the damping time scaled with the mass, also as a function of the compactness. In this case the empirical relation is given by
\begin{eqnarray}
\frac{10^3}{\tau(\mu s)} = 
\frac{1}{M(M_{\odot})}\left[a_2\left(\frac{M}{R_s}\right)^{2} + b_2\frac{M}{R_s} + c_2\right].
\label{empiricalomgtau}
\end{eqnarray} 
The fitting coefficients are also presented in  Table \ref{tab:universal fits} of the Appendix. In this case, the universality seems to be slightly broken at high values of the compactness, where the deviation in the lower panel can increase beyond $10^{-1}$. However, at least for the GR case, these configurations are beyond the stability region (i.e., they are beyond the maximum neutron star mass of each EOS), and we include them here only for completeness and comparison.

\begin{figure}
	\centering
		\includegraphics[width=.52\textwidth, angle =-90]{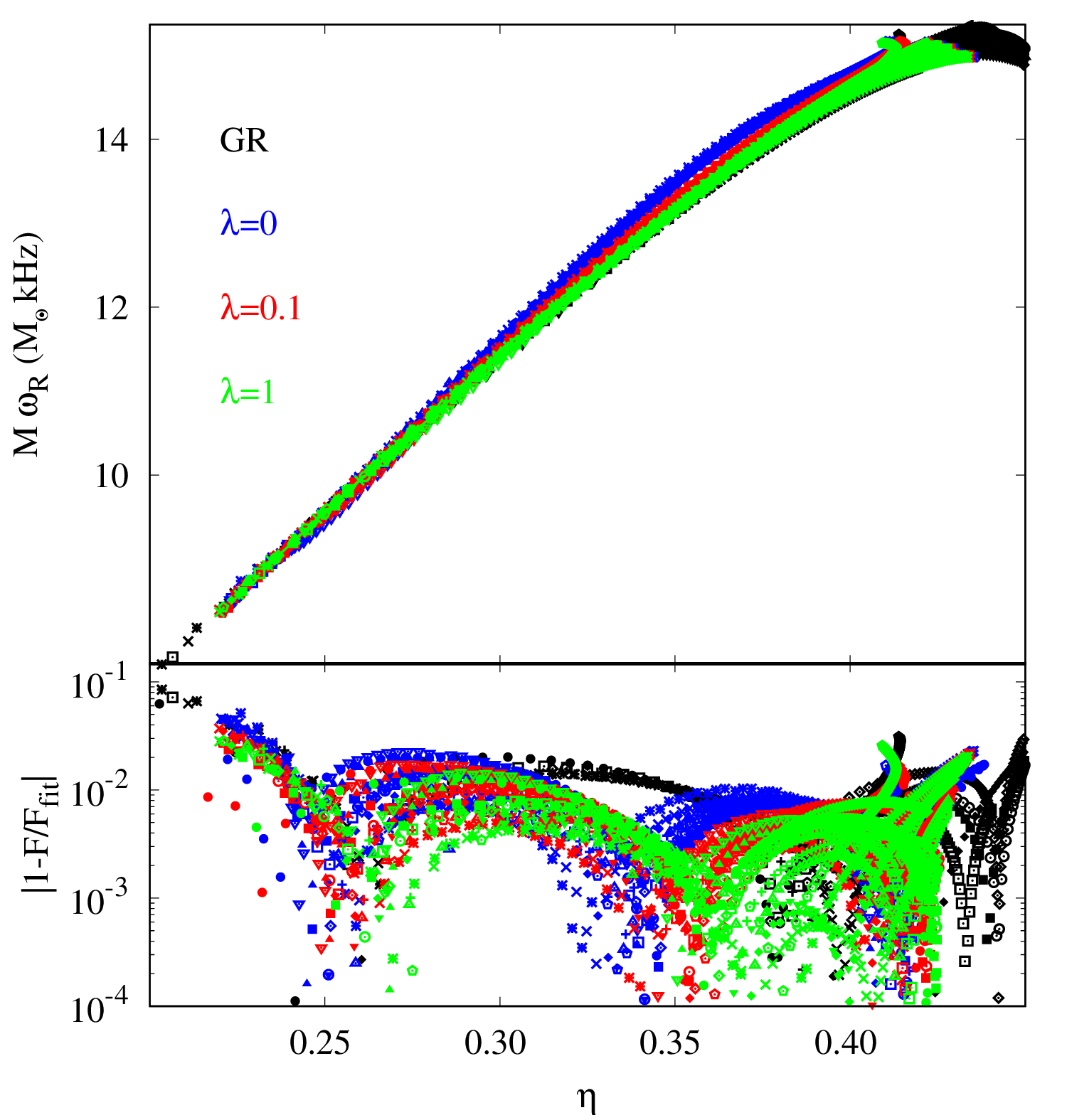}
		\includegraphics[width=.52\textwidth, angle =-90]{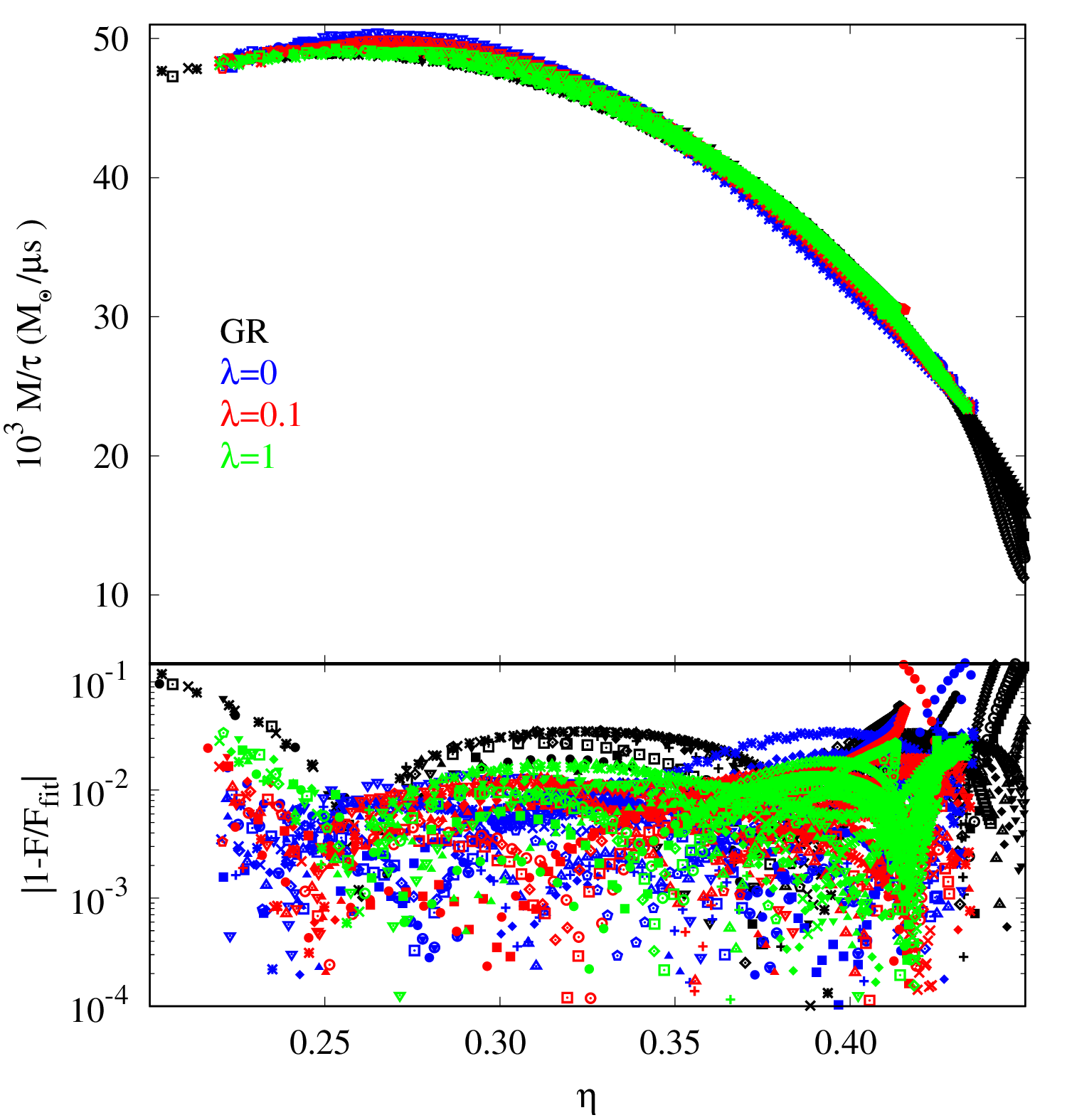}
	\caption{
		(left) Frequency $\omega_{R}$ in kHz scaled to total mass in $M_\odot$ as a function of the scaled moment of inertia $\eta=\sqrt{M^{3}/I}$ (fundamental $l=2$ wI mode). In black we show the GR configurations. In colors we show the data for scalarized stars with $A=e^{\frac{1}{2}\beta\varphi^2}$, $\beta=-6$, $m_{\varphi}=10^{-3}$ and different values of $\lambda$: in blue $\lambda=0$, in red $\lambda=0.1$ and in green $\lambda=0.1$. 
		The lower panel shows the relative difference $|1 - F/F_{\rm fit}|$ between the scaled data and the fitting relations presented in eq. (\ref{empiricalfrec2}) and Table \ref{tab:universal fits} . 
		(right) A similar figure but showing the inverse of the damping time $\tau$ in $\mu$s scaled to the total mass in $M_\odot$ as a function of the compactness $M/R_s$.
	}
	\label{plot_MR_omegaR_scaled_v3_w_error}
\end{figure}

In Fig.\ref{plot_MR_omegaR_scaled_v3_w_error} we show similar plots for the scaled frequency (left) and damping time (right), but this time as a function of the scaled moment of inertia $I$ of the neutron star, $\eta=\sqrt{M^{3}/I}$. The parameter $\eta$ can be understood as a generalized compactness, and it was shown in \cite{Lau2010} that it leads to universal relations that have a very good EOS independence. 

We can construct similar quadratic relations between the scaled frequency / damping time and the scaled moment of inertia,
\begin{eqnarray}
\omega(khz) = 
\frac{1}{M(M_\odot)}\left[a_3\eta^{2} + b_3 \eta + c_3\right], \label{empiricalfrec2}\\
\tau = 
\frac{1}{M}\left[a_4\eta^{2} + b_4 \eta + c_4\right].
\label{empiricalomgtau2}
\end{eqnarray}  
Making fits of the data with respect to these relations (the parameters are also shown in Table \ref{tab:universal fits} of the Appendix) reveals a similar picture as in the previous case where the standard compactness was used.
The  deviations of the data from the fitted relations are slightly smaller in the case of the frequency, especially for large values of the neutron star mass and compactness, but in general they are of the same order.

\begin{figure}
	\centering
		\includegraphics[width=.52\textwidth, angle =-90]{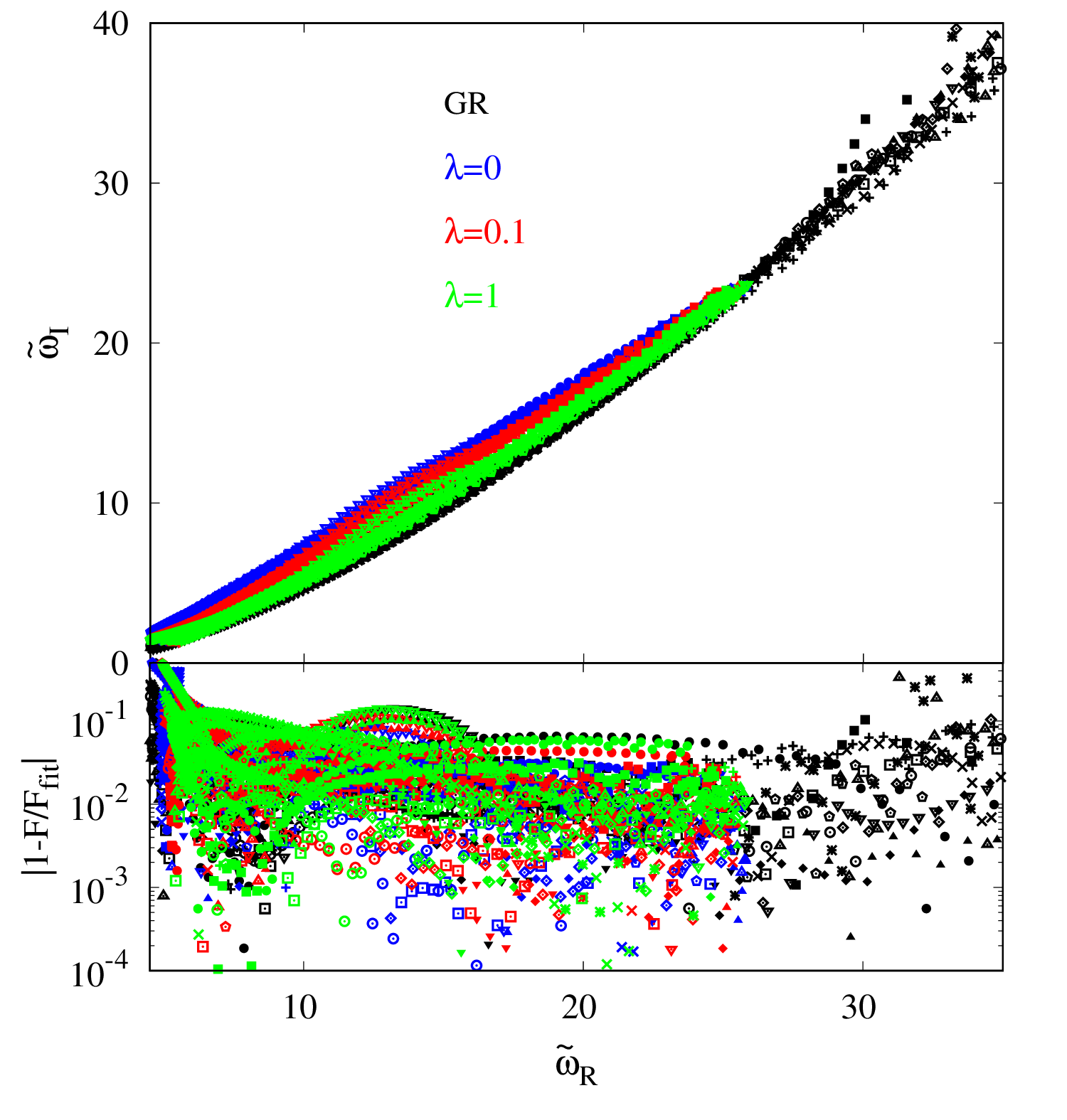}
		\includegraphics[trim=-2.5cm 0 0 -1cm,width=.48\textwidth, angle =-90]{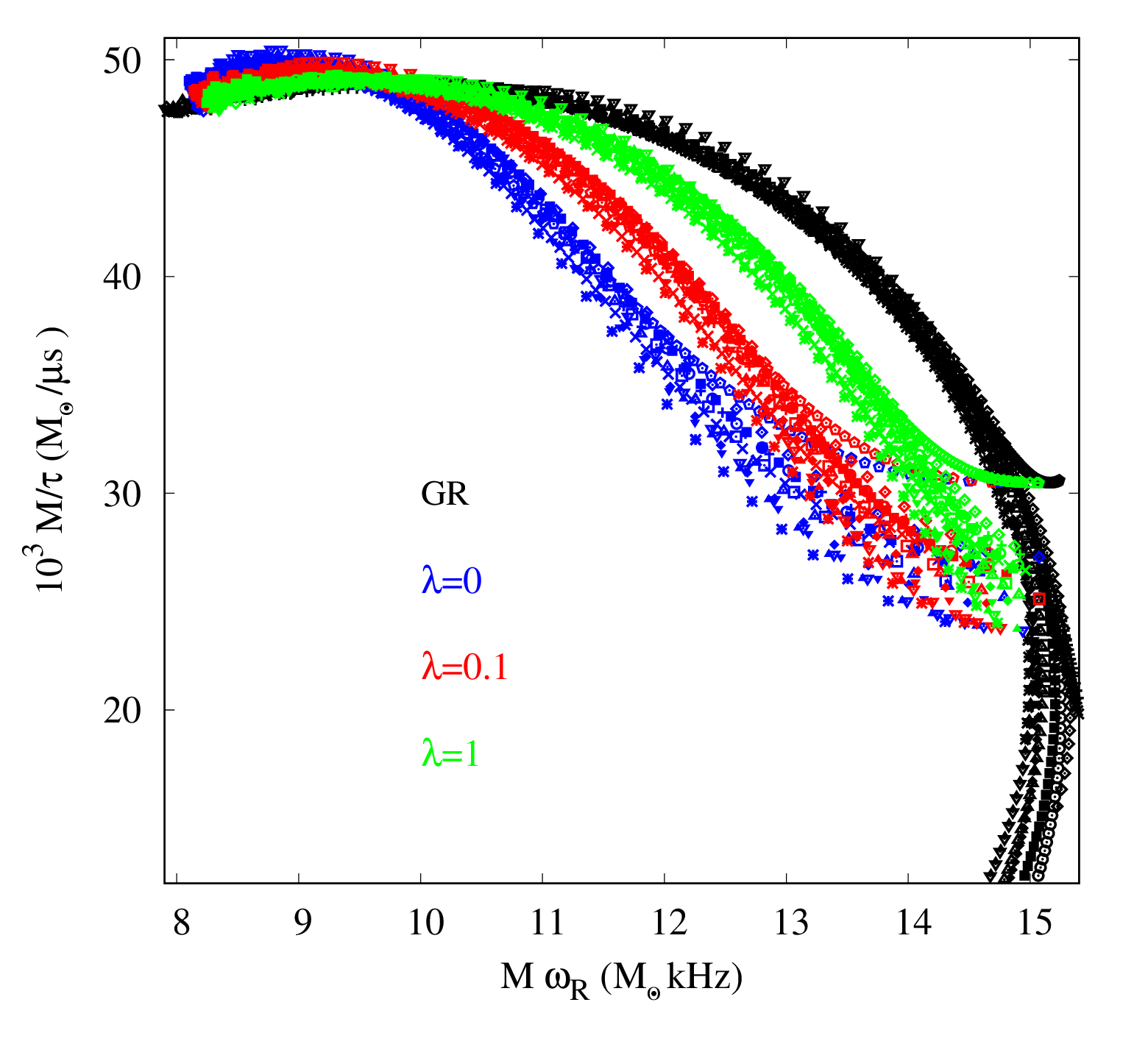}
	\caption{(left) The imaginary part of $\tilde\omega_I$ as a function of real part $\tilde\omega_R$, where $\tilde\omega_{R,I} = \frac{\omega_{R,I}}{\sqrt{p_0}}$ and $p_0$ is the central pressure. (right) Inverse of the damping time as a function of the frequency, both of them scaled to the total mass.
	}
	\label{plot_MomegaR_eta_scaled_error4}
\end{figure}

In Fig.\ref{plot_MomegaR_eta_scaled_error4}(left) we show the imaginary part of $\tilde\omega_I$ as a function of real part $\tilde\omega_R$, where $\tilde\omega_{R,I} = \frac{\omega_{R,I}}{\sqrt{p_0}}$ and $p_0$ is the central pressure. In this case we can fit the data to the phenomenological relation
\begin{eqnarray}
\tilde \omega_{I}=
a_5 \tilde \omega_{R}^{2}
+  b_5 \tilde \omega_{R}
+ c_5  
. \label{pc_scaled_modes}
\end{eqnarray}
The fitting parameters can be found in Table \ref{tab:universal fits}, as well. In the bottom panel of Fig.\ref{plot_MomegaR_eta_scaled_error4}(left) we see that the universality is in general of slightly less quality for small values of the mass (corresponding to small values of $\tilde\omega_R$).

In general, from the different universal relations we have considered, we learn that the scaled frequencies and damping times do not deviate significantly from GR for the different values of $\lambda$ we have considered. The deviation is always of the same order as the universality of the matter-independent empirical relations. In Fig.\ref{plot_MomegaR_eta_scaled_error4}(right) we show the inverse of the damping time as a function of the frequency, both of them scaled to the total mass. Here we can appreciate how the scalarized solutions present a qualitatively different behaviour from the GR configurations, specially for small or vanishing self-interaction. However, as we have seen with the previous fits, quantitatively this difference is very small, and pretty much of the same order as the variation with the matter composition. Hence the standard universal relations from GR describe to a large degree also the behaviour of the scalarized configurations. 
In principle, we could choose a different value of the coupling parameter $\beta$ in order to make this difference grow and the deviation from GR increase.

\section{Conclusions} \label{sec:Conclusions}

In this study, we have computed the axial QNMs of static neutron stars in massive STT with self-interaction. We have analyzed several realistic EOSs, including pure nuclear matter, hyperons and hybrid quarks+hyperons. We have focused on the $l=2$ fundamental wI mode.

We have compared standard GR configurations with scalarized stars with a particular form of the coupling function, $A=e^{\frac{1}{2}\beta\varphi^2}$, with $\beta=-6$ and $m_{\varphi}=10^{-3}$, values that are compatible with current observational constraints on this class of STT. We have analyzed in detail the impact of changing the values of the self-interaction parameter $\lambda$, considering models without self-interaction $\lambda=0$, and two cases with self-interaction $\lambda=0.1$ and $1$.

Regarding the spectrum, we have found that the scalarized stars tend to have lower frequencies than the unscalarized ones. The damping time on the contrary, does not seem to be very sensitive to the scalarization. This result is reminiscent of the one obtained in \cite{Blazquez-Salcedo2018}, where although the theory is different and there is no spontaneous scalarization, the neutron stars possess also a massive scalar field with self-interaction.

We have shown that increasing the value of the self-interaction makes the spectrum of QNMs become closer and closer to the spectrum of pure GR stars. The largest deviation occurs for models with $\lambda=0$. 

We have studied the impact of modifying the self-interaction parameter on several universal relations between the spectrum and the global quantities of the star (mass, radius, moment of inertia, etc). The results show that the impact of changing the self-interaction parameter is of the same order as changing the equation of state, at least for the particular values considered in this study. 

However, in principle, one could modify the parameter $\beta$ and/or the mass of the scalar field in order to increase the deviation of the scalarized NS properties from those of the GR ones. In this case the universal relations for the scalarized neutron stars could very well be quantitatively very different from the respective GR universal relations.
 
\section*{Acknowledgment}
We would like to acknowledge support by the
DFG Research Training Group 1620 {\sl Models of Gravity}
and the COST Actions CA16104 {\sl GWverse} and CA16214 {\sl PHAROS}.
JLBS would like to acknowledge support from the DFG project BL 1553. 
SY  would like to thank for the support by the Bulgarian NSF
Grant DCOST 01/6. DD would like to thank the European Social Fund, the
Ministry of Science, Research and the Arts Baden-Wurttemberg for the
support. DD is indebted to the Baden-W\"urttemberg Stiftung for the
financial support by the Eliteprogramme for Postdocs.

\section*{Appendix}

Here we present the fitting parameters $a_i$, $b_i$ and $c_i$, corresponding to the empirical relations of equations (\ref{empiricalfrec})-(\ref{pc_scaled_modes}). Each fit is done independently for each case: GR, $\lambda=0$, $\lambda=0.1$ and $\lambda=1$.


\begin{table}
	\centering
	\begin{tabular}{||c|c|c|c|||c|c|c|c||}
		\cline{2-4}\cline{6-8}
		\multicolumn{1}{c||}{\textbf{GR}} 
		& $\mathbf{a_i}$ & $\mathbf{b_i}$ & $\mathbf{c_i}$ 
 & \multicolumn{1}{c||}{$\mathbf{\lambda=0}$} 
& $\mathbf{a_i}$ & $\mathbf{b_i}$ & $\mathbf{c_i}$		
		\\ 
		\hline\hline 
		$i=\mathbf{1}$ & $-86 \pm 2$ & $79 \pm 1$ & $-0.036 \pm 0.09$ 
		&$i=\mathbf{1}$ & $16 \pm 4$ & $32 \pm 2$ & $4.2 \pm 0.2$
		\\ 
		\hline 
		$i=\mathbf{2}$ & $-1775 \pm 34$ & $548 \pm 13$ & $7 \pm 1$
        &$i=\mathbf{2}$ & $-1294 \pm 39$ & $440 \pm 16$ & $13 \pm 2$
		 \\ 
		\hline 
		$i=\mathbf{3}$ & $-84\pm2$ & $88\pm1$ & $-7.3\pm0.2$
		&$i=\mathbf{3}$ & $-91\pm1$ & $93\pm1$ & $-8.2\pm0.1$
		 \\ 
		\hline 
		$i=\mathbf{4}$ &$-1143 \pm 12$ & $640 \pm 10$ & $-40 \pm 2$
		&$i=\mathbf{4}$ &$-919 \pm 4$ & $485 \pm 2$ & $-14.0 \pm 0.4$
		 \\ 
		\hline 
		$i=\mathbf{5}$ & $0.0211 \pm 0.0005$ & $0.44 \pm 0.02$ & $-1.5 \pm 0.1$
		&$i=\mathbf{5}$ & $-0.0011 \pm 0.0003$ & $1.08 \pm 0.01$ & $-3.61 \pm 0.05$
		 \\ 
		\hline 
		\hline
		\cline{2-4}	\cline{6-8}
\multicolumn{1}{c||}{$\mathbf{\lambda=0.1}$} 
& $\mathbf{a_i}$ & $\mathbf{b_i}$ & $\mathbf{c_i}$
&\multicolumn{1}{c||}{$\mathbf{\lambda=1}$} 
& $\mathbf{a_i}$ & $\mathbf{b_i}$ & $\mathbf{c_i}$ \\ 
\hline\hline \hline 
$i=\mathbf{1}$ & $-28 \pm 3$ & $54 \pm 1$ & $2.2 \pm 0.1$ 
&$i=\mathbf{1}$ & $-83 \pm 3$ & $77 \pm 1$ & $0.13 \pm 0.13$\\ 
\hline 
$i=\mathbf{2}$ & $-1427 \pm 27$ & $463 \pm 10$ & $12 \pm 1$ 
&$i=\mathbf{2}$ & $-1374 \pm 20$ & $419 \pm 8$ & $17 \pm 1$\\ 
\hline 
$i=\mathbf{3}$ & $-72\pm1$ & $81\pm1$ & $-6.2\pm0.1$
&$i=\mathbf{3}$ & $-63\pm1$ & $74\pm1$ & $-5.0\pm0.1$ \\ 
\hline 
$i=\mathbf{4}$ & $-918 \pm 4$ & $488 \pm 3$ & $-15.4 \pm 0.4$ 
&$i=\mathbf{4}$ & $-941 \pm 4$ & $508 \pm 3$ & $-19.5 \pm 0.5$ \\ 
\hline
$i=\mathbf{5}$ & $0.0040 \pm 0.0002$ & $0.96 \pm 0.01$ & $-3.65 \pm 0.05$ 
&$i=\mathbf{5}$ & $0.0151 \pm 0.0003$ & $0.63 \pm 0.01$ & $-2.36 \pm 0.05$\\ 
\hline 		
		\hline 
	\end{tabular} 
	\caption{Parameters of the fits for the empirical relations of equations (\ref{empiricalfrec})-(\ref{pc_scaled_modes}), including all realistic EOSs.}
	\label{tab:universal fits}
\end{table}

\bibliographystyle{ieeetr}
\bibliography{references}

\end{document}